\documentclass[letterpaper,twocolumn,accepted=2026-07-14]{quantumarticle} 
\pdfoutput=1
\usepackage{amsmath,amsthm,amsfonts,amssymb}
\usepackage{graphicx}
\usepackage{dcolumn}
\usepackage{bm}
\usepackage[table]{xcolor}

\usepackage[numbers]{natbib}
\usepackage{url}

\usepackage{csquotes}

\usepackage{multirow,makecell,hhline,array}

\usepackage{tikz}
\usetikzlibrary{quantikz}
\usetikzlibrary{positioning,arrows,shapes,shadows,calc,fit,backgrounds}

\usepackage[ruled,vlined]{algorithm2e}

\newcommand{\CC}{\mathbb{C}} 

\newcommand{\cB}{\mathcal{B}}

\newcommand{\CA}{\mathbf{C}_{\mathrm{amp}}}
\newcommand{\sCA}{\widehat{\mathbf{C}}_{\mathrm{amp}}}
\newcommand{\CS}{\mathbf{C}_{\mathrm{samp}}}
\newcommand{\sCS}{\widehat{\mathbf{C}}_{\mathrm{samp}}}
\newcommand{\qsize}{\mathrm{qsize}}

\usepackage[caption=false]{subfig}

\usepackage{xurl}

\usepackage[unicode,colorlinks=true,
             linkcolor=blue,citecolor=blue,urlcolor=blue]{hyperref}

\newcommand{\abs}[1]{{\lvert#1\rvert}}

\newtheorem{proposition}{Proposition}

\begin{document}

\title{Pilot-Wave Simulator: Exact Classical Sampling from
Ideal and Noisy Quantum Circuits up to Hundreds of Qubits} 

\author{Gleb Kalachev}%
\affiliation{Lomonosov Moscow State University, Moscow, 119991, Russia}
\orcid{0000-0003-2695-3179}

\author{Pavel Mosharev}
\email{mosharev@ustc.edu.cn}
\orcid{0000-0001-7065-1974}
\affiliation{School of Physical Sciences, University of Science and Technology of China, Hefei, 230026, China}

\author{Zuoheng Zou}%
\orcid{0000-0001-5664-0179}
\affiliation{School of Physics, Peking University, Beijing 100871, China}%

\author{Pavel Panteleev}%
\affiliation{Lomonosov Moscow State University, Moscow, 119991, Russia}
\orcid{0000-0002-6772-7867}

\author{Man-Hong Yung}%
\orcid{0000-0001-7376-435X}
%
\affiliation{Shenzhen Institute for Quantum Science and Engineering, Southern University of Science and Technology, Shenzhen 518055, China}%


\begin{abstract}
Quantum circuit simulators running on classical computers offer a vital platform for designing, testing, and optimizing quantum algorithms, driving innovation despite limited access to real quantum hardware. However, their scalability is inherently constrained by exponential memory and computational overhead, which restricts accurate simulation of large-scale quantum circuits and often results in approximate output distributions. Here, we propose an exact sampling algorithm that integrates tensor network contraction techniques with a Markov process, wherein a classical state evolves according to the local structure of the quantum circuit. As a demonstration, we target the challenge of generating samples from ideal and noisy QAOA circuits with up to 476 qubits, incorporating both depolarizing and amplitude damping noise models. These results provide exact-sampling data for structured QAOA circuits in a regime where full-state simulation is not practical, and give a benchmark for selected local noise models.
\end{abstract}

\maketitle

\section{Introduction}

In recent years, classical simulation of quantum circuits has become a~key tool both for verifying the behavior of NISQ-era devices and for exploring the potential of hybrid quantum-classical algorithmic ideas. These simulations vary in scope and computational approaches. The most straightforward approach, called \emph{state-vector} simulation, computes the amplitudes of all possible quantum states of a circuit \cite{Wu2019, Zhang2023, Gangapuram2024}. An~analogous approach for noisy circuits is called \emph{density-matrix} simulation \cite{OBrien2017, Li2021}. Both of these methods are very expensive and cannot be used to simulate more than a~few dozen qubits even on a~supercomputer. Matrix-product state (MPS) method can simulate circuits up to a~few hundred qubits when entanglement remains low at the cost of limited precision \cite{Vidal2003, Zhou2020, Ayral2023, dubey2025}. Other methods such as gPEPS~\cite{Patra2024} or Pauli Path Simulator (PPS) \cite{GonzalezGarcia2025paulipath} aim to estimate expectation values of observables rather than to produce samples.

Currently, many simulators for quantum circuits are based on tensor network methods. These tools allow efficient sampling from the output of \emph{random} quantum circuits using \emph{rejection sampling} methods \cite{Villalonga:2019}. However, they are not well-suited for producing samples from the outputs of general \emph{non-random} circuits, where output probabilities deviate significantly from uniformity, and one has to evaluate many amplitudes to produce just one sample.

An alternative approach, called \textsc{SEQCSim}, was proposed earlier in~\cite{pilot-wave-sim-2}. In this framework, the evolution of the quantum state is accompanied by the stochastic evolution of a~classical configuration, enabling the generation of samples with perfect fidelity while accessing only a~very small subset of amplitudes at each step.

A similar method was introduced by Bravyi, Gosset, and Liu in their \emph{gate-by-gate} simulation algorithm \cite{bravyiHowSimulateQuantum2022}, which provides a~rigorous framework for simulating quantum measurements without computing marginal distributions. Although the original motivation in~\cite{bravyiHowSimulateQuantum2022} is different, the proposed gate-by-gate algorithm turns out to be  essentially the same as in the sequential simulation from~\cite{pilot-wave-sim-2}: a Markov process where a~classical state evolves based on the local structure of the quantum circuit.

In this work, we build on these ideas and propose a~new algorithm that combines the simulation frameworks from~\cite{pilot-wave-sim-2, bravyiHowSimulateQuantum2022} with modern tensor network contraction techniques~\cite{MultiampSim}. This hybrid approach allows efficient sampling from structured, non-random quantum circuits (including QAOA circuits with realistic noise models) by computing only a~small number of relevant amplitudes per gate. Unlike rejection sampling methods \cite{Villalonga:2019}, which are effective only when output probabilities are nearly uniform, our approach remains viable in highly non-uniform settings typical for optimization circuits.

Our simulator calculates the samples by sequentially updating the classical state according to transition probabilities that drive the evolution. As was pointed out in~\cite{pilot-wave-sim-2}, this process resembles Bohm's \emph{pilot-wave} interpretation of quantum mechanics~\cite{Bohm1952, pilot-wave-theories} where hidden deterministic evolution is driven by a nonlocal wave function.  For this reason, we refer to our algorithm as the \emph{Pilot-Wave} (PW) simulator.

The approaches~\cite{pilot-wave-sim-2, bravyiHowSimulateQuantum2022} share the same Markovian sampling principle: a~classical bitstring is updated after each gate according to local transition rules so that the instantaneous distribution matches the circuit’s measurement distribution. \textsc{SEQCSim}~\cite{pilot-wave-sim-2} introduced this paradigm and implemented it with a Feynman-path amplitude engine using polynomial memory, together with practical engineering optimizations. The gate-by-gate framework of Bravyi, Gosset, and Liu~\cite{bravyiHowSimulateQuantum2022} provides a~rigorous analysis of such Markovian updates and is agnostic to the backend used to compute the required amplitudes (e.g., full-state, path-sum, or tensor network methods).

There are also several software implementations of the gate-by-gate sampling idea. Shapiro and LaRose introduced \textsc{BGLS}, a~Python
package implementing the gate-by-gate sampling algorithm of Bravyi, Gosset, and Liu~\cite{ShapiroLaRose2023BGLS}. In addition, the \texttt{quimb} tensor-network package~\cite{GrayKourtis2021Cotengra, Gray2018Quimb} contains a~\texttt{sample\_gate\_by\_gate} routine. These implementations use the same general sampling principle, but differ in their amplitude backends, caching strategies, and practical optimizations.

Our contribution is a~version of this method designed for structured, non-random circuits: (i) we use a~tensor network amplitude oracle~\cite{MultiampSim} that evaluates only the few amplitudes needed at each step; (ii) we restrict stochastic branching via an~explicit block decomposition of each gate matrix, which bounds the number of candidate updates per gate; and (iii) we treat (sub)monomial (i.e., permutation–diagonal) gates deterministically, avoiding amplitude calls. Moreover, this deterministic treatment allows us to extend our algorithm to realistic noise models (e.g., Pauli/dephasing) with nearly the same time complexity as in the noiseless case. 

We emphasize that our method is not a~new tensor-network contraction algorithm and does not reduce the cost of evaluating a~single amplitude.
The tensor network is used as an~amplitude oracle. Moreover, the low-overhead noisy simulation described here applies to noise
channels that admit a~\emph{local-submonomial} Kraus representation, in the sense made precise in Appendix~\ref{app:noise-sim}. This includes the Pauli, depolarizing, phase-damping, and amplitude-damping channels used in our experiments. However, it does not cover general coherent calibration errors or general coherent entangling errors. Such errors can still be represented within the general Kraus framework, but they may change the tensor-network structure and increase the contraction complexity.

We put our simulator to the test on a~set of non-random quantum circuits. We use depth‑$p\le 3$ QAOA instances for the Ising model on grid‑like graphs to demonstrate the exponential decrease in ground state probability that has been conjectured in~\cite{augustino2024}, and shown in simulations in~\cite{Montanez-Barrera2025} and~\cite{Willsch2022} for a smaller number of qubits. We extend these original results from $N=28$ and $N=40$ to $N=49$ qubits, producing up to $\approx 10^{9}$ samples per circuit. We show that the distribution of samples reproduces the “pseudo‑Boltzmann’’ energy profile reported in Refs.~\cite{Diez-Valle23, diezvalle2025, Diez-Valle24}. We extend the original result from $N=24$ to $N=36$ qubits and show that for circuit depth $p \le 3$ the low-energy tail follows the “pseudo‑Boltzmann’’ trend with effective temperature decreasing as circuit depth grows.  

Because practical devices are inevitably noisy, we re‑run the same study under Pauli‑error and amplitude‑damping channels chosen to be representative of early‐generation superconducting processors.  Noise raises the effective temperature and suppresses low‑energy tails quite predictably, yet the simulator remains fast enough to deliver statistically significant samples for grids of up to $49$ qubits, providing a~quantitative baseline for near‑term hardware.

\begin{table}[!t]
        \setlength{\tabcolsep}{4pt}
        \renewcommand{\arraystretch}{1.1}
	\begin{tabular}{|l|c|c|c|c|}
        \hline
		{\sffamily\bfseries GS probability scaling}             & \#qubits   & $p$   & ideal & noisy \\ \hline
		Full-state \cite{augustino2024}      & 28  & 11  & $\checkmark$     & ---     \\ \hline
		Full-state \cite{Montanez-Barrera2025} & 35  & 200 & $\checkmark$     & ---     \\ \hline
		Full-state \cite{Willsch2022} & 40  & 13 & $\checkmark$     & ---     \\ \hline
		\textrm{Pilot-Wave}                                    & 49  & 3   & $\checkmark$     & $\checkmark$     \\ \hline
        \hline
		{\sffamily\bfseries Probability distribution}                & \#qubits   & $p$   & ideal & noisy \\ \hline
		Full-state \cite{Diez-Valle23}         & 24  & 1   & $\checkmark$     & ---     \\ \hline
		Full-state \cite{diezvalle2025}        & 16  & 15  & $\checkmark$     & ---     \\ \hline
		\textrm{Pilot-Wave}                                    & 36  & 3   & $\checkmark$     & $\checkmark$     \\ \hline
        \hline
		{\sffamily\bfseries Large-scale simulation}                & \#qubits   & $p$   & ideal & noisy \\ \hline
		\begin{tabular}[c]{@{}l@{}} Tensor network \cite{Lykov2022} \\ (single amplitude)\end{tabular} & 210 & 1   & $\checkmark$     & ---     \\ \hline
		\begin{tabular}[c]{@{}l@{}}Full-state \cite{Montanez-Barrera2025} \\ (sampling)\end{tabular}   & 42  & 400 & $\checkmark$    & ---     \\ \hline
		\begin{tabular}[c]{@{}l@{}} \textrm{Pilot-Wave} 
        \end{tabular}                   & 476 & 1   & $\checkmark$     & $\checkmark$     \\ \hline
	\end{tabular}
        \caption{Comparison of previously reported QAOA simulations. \emph{GS probability scaling} describes how the ground-state success probability scales with qubit number and circuit depth; \emph{Probability distribution} refers to analyses of full output distributions; \emph{Large-scale simulation} covers largest-qubit results using tensor-network or sampling approaches. Our Pilot-Wave algorithm enables exact sampling at larger qubit counts (ideal and noisy) and provides sampling-based estimates of output distributions rather than full distributions.}
\end{table}

Finally, to gauge whether shallow‑depth QAOA really offers an~advantage over purely classical heuristics, we experimentally support the conjecture made by Hastings in~\cite{hastings2019} that a classical local update rule proposed in the same work would have performance essentially similar to QAOA at the same depth. We evaluate it on grid Ising models of $100$ and $256$ qubits, demonstrating that the local update algorithm matches the performance of the ideal QAOA circuits at depths $1$ and $2$, while in the noisy setting the classical update already outperforms QAOA (Fig.~\ref{fig:Hastings_and_QAOA}).

We also show in Section~\ref{app:times} how the time to obtain a~small batch scales with the circuit size up to $N=476$ qubits for QAOA circuits of depth $p = 1$. Previous exact sampling results for QAOA were obtained only for $42$ qubit circuit in full-state simulation on a cluster of 3744 GPUs \cite{Montanez-Barrera2025}. Single amplitude simulation was performed in \cite{Lykov2022} for $N=210$ qubits at $p=1$ using the tensor network method on $1024$ nodes of supercomputer Theta. There are also simulations that compute expectation values of the energy for QAOA on $N=512$ qubits at $p \le 5$ \cite{augustino2024}, and approximate sampling for $N=256$ qubits at $p = 4$~\cite{dubey2025}.
The experiments in our work are performed on two servers that have 512 and 96 CPU cores, respectively, which demonstrate the practical scope of the approach.

\section{Pilot-Wave simulator}
\label{sec:PW_sim}

Suppose we have a~discrete-time unitary evolution on an~$n$-qubit system. The main idea of the \emph{Pilot-Wave} (PW) simulator for this system is to introduce a~Markov chain with the state space $\{0,1\}^n$, where the classical state $s_t \in \{0,1\}^n$ evolves stochastically according to the wave-function $\psi_t \in \CC^{2^n}$ playing the role of the pilot-wave, i.e., we have 
\[\Pr(s_t = x) = \abs{\braket{x}{ \psi_t}}^2\] 
for all $x\in \{0,1\}^n$. Furthermore, we also assume that our discrete-time unitary evolution satisfies a~certain \emph{locality condition}: at each moment $t>0$, the transition probabilities
\[p_t(x \to y) := \Pr(s_{t} = y \mid s_{t-1} = x)\]
from the current~state $x$ to the next state $y$ depend only on $L = O(1)$ amplitudes from the wave-function $\psi_{t}$. Thus to be able to sample from the probability distribution 
\[p_t(x) := \Pr(s_t = x) = \abs{\braket{x}{ \psi_t}}^2\]
at each time $t$ using our Markov chain, we need to keep track of the classical state $s_t\in \{0,1\}^n$ and the $L$ amplitudes of the quantum state $\psi_{t}\in\CC^{2^n}$. This significantly reduces the memory size since we do not need the full quantum state $\psi_{t}$ to sample from it.  However, we still need to find the $L$ amplitudes from $\psi_{t}$ in a~computationally efficient way. To achieve this goal we propose to use tensor networks.

As we will see below, the locality condition required to use the proposed PW simulator can always be satisfied if our discrete-time  unitary evolution is described using the quantum circuit model. 
Let us recall that a~\emph{quantum circuit} can be viewed as a~sequence of unitary matrices $U_1,...,U_T$ called \emph{gates}, where the gate $U_t$ is applied at time $t$, and it affects only a~\emph{constant} number $m$ of qubits from the $n$-qubit state $\psi_t \in \CC^{2^n}$. In this case, we define our~discrete-time unitary evolution in the following way: $\psi_0=\ket{0}^{\otimes n}$, $\psi_t=U_t\ket{\psi_{t-1}}$, $t\in \{1,\dots,T\}$. Below, we assume that all $2^n \times 2^n$ matrices $U_t$ are given in the computational basis $\cB=\{\ket{x}\mid x\in\{0,1\}^n\}$.

Let $p_t(x)=|\braket{x}{\psi_t}|^2$. Our aim is to define a~Markov chain with transition probabilities $p_t(x\to y)$ in such a~way that: 
\begin{equation}\label{eqn:p-i2j}
p_{t}(y)=\sum_{x\in \{0,1\}^n}p_{t-1}(x)p_t(x\to y).
\end{equation} 

Consider two cases. The first case is when the unitary matrix $U_t$ is \emph{monomial}, i.e., $U_t = M_\sigma D$ where $M_\sigma$ is a~permutation matrix for some permutation $\sigma \in \mathbf{S}_{2^n}$ and $D$ is a~diagonal matrix. In this case, $U_t$ acts as a~composition of the permutation $\sigma$ and a~phase shift on the state-vector $\psi_{t-1}$. Since phase shifts do not change probabilities, we can define the transition probability as follows:
$$p_t(x\to y)=\begin{cases} 
  1&\mbox{if }y=\sigma(x),\\
  0&\mbox{else}.
\end{cases}$$
It is easy to see that condition \eqref{eqn:p-i2j} is satisfied in this case.

Consider the case of a~non-monomial matrix. Decompose each matrix 
$U_t=\bigoplus_{k=1}^{m_t}U^{(k)}_t$, where $U^{(k)}_t$ acts on subspace $V_t^{(k)}$ with basis $\cB_t^{(k)}\subseteq \cB$ (see Fig. \ref{fig:decompose}). Since $\bigsqcup_{k=1}^{m_t}\cB_t^{(k)}=\cB$, for each vector $\ket{x}\in \cB$ there is an index $k_t(x)$ such that $\ket{x}\in \cB_t^{(k_t(x))}$.

Let $p_t(x)=|\braket{x}{\psi_t}|^2$. Let us define the transition probabilities 
\[
  p_t(x\to y)=\begin{cases} 
  \displaystyle{\frac{p_{t}(y)}{\sum_{l\in \cB_t^{(k_t(y))}}p_{t}(l)}}&\mbox{if }k_{t}(x)=k_{t}(y),\\
  0&\mbox{else},
\end{cases}
\]
where, on zero-mass blocks, $p_t(x\to y)=\mathbf{1}\{y=x\}$ by definition; such states are never visited.

It is clear that we have 
$$\sum_{y\in\{0,1\}^n}p_t(x\to y)=1.$$
Let us check condition \eqref{eqn:p-i2j}.
Since $U_t^{(k)}$ is unitary and $\psi_t|_{V_t^{(k)}}=U_t^{(k)}(\psi_{t-1}|_{V_t^{(k)}})$, we have:

\begin{align*}
\sum_{\ket{x}\in \cB_t^{(k)}} p_{t-1}(x)
   &= \|\psi_{t-1}|_{V_t^{(k)}}\|^2
    = \|\psi_t|_{V_t^{(k)}}\|^2
\\
   &= \sum_{\ket{x}\in \cB_t^{(k)}} p_t(x),
\end{align*}

\begin{align*}
\sum_{x\in \{0,1\}^n} p_{t-1}(x) p_t(x\!\to\! y)
  &= \sum_{\ket{x}\in \cB_t^{(k_t(y))}}
     \frac{p_{t-1}(x)p_t(y)}{\displaystyle
       \sum_{\ket{l}\in \cB_t^{(k_t(y))}} p_t(l)}
\\[4pt]
  &= p_t(y).
\end{align*}

Thus, \eqref{eqn:p-i2j} holds in both cases.
Therefore, we can define the Markov chain $s_t\in\{0,1\}^n$, $s_0=\mathbf{0}$, $\Pr (s_{t+1} = y\mid s_{t}=x)=p_t(x\to y)$.
Condition \eqref{eqn:p-i2j} guarantees that $\Pr (s_{t}=x)=p_t(x)$. 

\begin{algorithm}[ht]
  \SetKwProg{Function}{Function}{:}{}
  \SetKwFunction{Sample}{Sample}
  \SetKwFunction{UpdateState}{UpdateState}
  \SetKwFunction{BlockDecompose}{BlockDecompose}

  \Function{\Sample{$C$}}{
     $n \gets$ number of qubits in $C$\;
     $s \gets \mathbf{0}\in\{0,1\}^n$\;
     \For{$t:=1$ \KwTo {\rm number of gates in} $C$}{
        $C_t\gets$ subcircuit containing first $t$ gates of $C$\;
        $s \gets $ \UpdateState{$C_t, s$}\;
     }
     \Return{s}\;
  }
  \Function{\UpdateState{$C,s$}}{
    Let $A$ be the matrix of the last gate of circuit $C$ acting on qubit set $q$\;
    \eIf{$A$ {\rm represents a~composition of a~permutation $\sigma$ and a~diagonal matrix on }$\{0,1\}^{|q|}$}{
        $s[q]\gets \sigma(s[q])$
    }{
        $b \gets$ \BlockDecompose{$A,s,q$}\;
        \eIf{$|b|>1$}{
            Calculate $p_i:=|\mathbf{amp}(C,i)|^2$ for $i\in b$\;
            $p_{\mathrm{block}}\gets\sum_{i\in b}p_i$\;
            Choose $s'$ from $b$ according to the distribution $\{p_i/p_{\mathrm{block}}\}_{i\in b}$\;
            $s\gets s'$\;
        }{
            $s\gets $ (unique) element of $b$\;
        }
    }
    \Return{$s$}\;
  }
  \Function{\BlockDecompose{$A,s,q$}}{
     Build graph $G=(\{0,1\}^{|q|},E)$ where $E=\{\{x,y\}\mid A_{x,y}\ne 0\}$\;
     Find connected component $r$ of $G$ containing $s[q]$\;
     \Return{$\{x\in\{0,1\}^n:\ x[\bar q]=s[\bar q]\mbox{ and }x[q]\in V(r)\}$}\;
  }
\caption{Procedure $\mathbf{sample}(C)$.}
\label{al:sample}
\end{algorithm}

\begin{figure*}[ht]
    \newcommand{\zzero}[1]{\node[black!30!white] at (#1) {0}}
    \newcommand{\lbr}[1]{\draw[thick] (#1,-1.8) -- ++(-0.1,0) -- ++(0,2.1) -- ++(0.1,0);}
    \newcommand{\rbr}[1]{\draw[thick] (#1,-1.8) -- ++(0.1,0) -- ++(0,2.1) -- ++(-0.1,0);}
\subfloat[Matrix $A\otimes I_{2^{n-m}}$]{\label{fig:block-prod}
    \begin{tikzpicture}
        \node at (0,0) {$A$};
        \zzero{0,-0.7};
        \node at (0,-1.4) {$\vdots$};
        \zzero{0,-2.1};
        
        \zzero{0.7,0};
        \node at (0.7,-0.7) {$A$};
        \node at (0.7,-1.4) {$\vdots$};
        \zzero{0.7,-2.1};
        
        \node at (1.4,0) {$\cdots$};
        \node at (1.4,-0.7) {$\cdots$};
        \node at (1.4,-1.4) {$\ddots$};
        \node at (1.4,-2.1) {$\cdots$};
        
        \zzero{2.1,0};
        \zzero{2.1,-0.7};
        \node at (2.1,-1.4) {$\vdots$};
        \node at (2.1,-2.1) {$A$};

        \draw (1.75,-1.75) rectangle (2.45,-2.45) rectangle (-0.35,0.35) rectangle (0.35,-0.35) rectangle (1.05,-1.05);
    \end{tikzpicture}
    }
    \hspace{0.5cm}
\subfloat[Monomial gates]{\label{fig:simple-gate}
    \begin{tikzpicture}
        \lbr{-0.2}

        \node at (0,0) {1};
        \zzero{0,-0.5};
        \zzero{0,-1};
        \zzero{0,-1.5};
        
        \zzero{0.5,0};
        \node at (0.5,-0.5) {1};
        \zzero{0.5,-1};
        \zzero{0.5,-1.5};
        
        \zzero{1,0};
        \zzero{1,-0.5};
        \zzero{1,-1};
        \node at (1,-1.5) {1};
        
        \zzero{1.5,0};
        \zzero{1.5,-0.5};
        \node at (1.5,-1) {1};
        \zzero{1.5,-1.5};
        
        \rbr{1.7}
        
        \node at (0.75, -2.25) {CNOT};
    \end{tikzpicture}
    \hspace{0.3cm}
    \begin{tikzpicture}
    \lbr{-0.2} \rbr{1.7}
        \node at (0,0) {1};
        \zzero{0,-0.5};
        \zzero{0,-1};
        \zzero{0,-1.5};
        
        \zzero{0.5,0};
        \zzero{0.5,-0.5};
        \node at (0.5,-1) {$i$};
        \zzero{0.5,-1.5};
        
        \zzero{1,0};
        \node at (1,-0.5) {$i$};
        \zzero{1,-1};
        \zzero{1,-1.5};
        
        \zzero{1.5,0};
        \zzero{1.5,-0.5};
        \zzero{1.5,-1};
        \node at (1.5,-1.5) {1};
        \node at (0.75, -2.25) {iSWAP};
    \end{tikzpicture}
    }
    \hspace{0.5cm}
    \subfloat[General gates]{ \label{fig:block-gate}
    \begin{tikzpicture}
    \lbr{-0.3} \rbr{3.15}
        \node at (0,0) {1};
        \zzero{0,-0.5};
        \zzero{0,-1};
        \zzero{0,-1.5};
        
        \zzero{0.5,0};
        \node at (0.5,-0.5) {1};
        \zzero{0.5,-1};
        \zzero{0.5,-1.5};
        
        \zzero{1.3,0};
        \zzero{1.3,-0.5};
        \node at (1.3,-1) {$\cos \frac\theta 2$};
        \node at (1.3,-1.5) {$-i\sin\frac\theta 2$};
        
        \zzero{2.5,0};
        \zzero{2.5,-0.5};
        \node at (2.5,-1) {$-i\sin\frac\theta 2$};
        \node at (2.5,-1.5) {$\cos\frac\theta 2$};

        \draw (0.25, -0.25) rectangle (-0.25,0.25);
        \draw (0.25, -0.25) rectangle (0.75, -0.75);
        \draw (3.1, -1.75) rectangle (0.75, -0.75);

        \node at (1.25, -2.25) {controlled $RX(\theta)$};
    \end{tikzpicture}
    \hspace{0.3cm}
    \begin{tikzpicture}
    \lbr{-0.6} \rbr{2.95}
        \node at (-0.3,0) {1};
        \zzero{-0.3,-0.5};
        \zzero{-0.3,-1};
        \zzero{-0.3,-1.5};
        
        \zzero{0.5,0};
        \node at (0.5,-0.5) {$\cos\theta$};
        \node at (0.5,-1) {$-i\sin\theta$};
        \zzero{0.5,-1.5};
        
        \zzero{1.5,0};
        \node at (1.5,-0.5) {$-i\sin\theta$};
        \node at (1.5,-1) {$\cos\theta$};
        \zzero{1.5,-1.5};
        
        \zzero{2.5,0};
        \zzero{2.5,-0.5};
        \zzero{2.5,-1};
        \node at (2.5,-1.5) {$e^{i\phi}$};

        \draw (-0.05, -0.25) rectangle (-0.55,0.25);
        \draw (-0.05, -0.25) rectangle (2.1, -1.25);
        \draw (2.9, -1.75) rectangle (2.1, -1.25);
        
        \node at (0.75, -2.25) {$\mathrm{fSim}(\theta,\phi)$};
    \end{tikzpicture}
    }
    \caption{Panel (a) shows that $U_t$ has the block structure of $A \otimes I_{2^{n-m}}$ (up to a~basis permutation).
Each block corresponds to the gate matrix $A$, which can be either (b) a~monomial gate (i.e., exactly one nonzero per row and column)
 or (c) a~general gate. Matrices of general gates are
further partitioned into blocks using the \textsc{BlockDecompose} procedure from
Algorithm~\ref{al:sample}.}
\label{fig:decompose}
\end{figure*}

To generate a~sample we should sequentially calculate $s_1,\dots,s_T$ as shown in Algorithm~\ref{al:sample}. Here for a~bit string $x\in\{0,1\}^n$ and an index set $q\subseteq \{1,\dots,n\}$, $x[q]$ denotes the substring on $q$ and $x[\bar q]$ the substring on its complement $\bar q=\{1,\dots,n\}\setminus q$.
At each step, we  calculate $|\cB_t^{(k)}|$ probabilities, where for an~$n$-qubit circuit $C_t = U_t \dots U_1$ and a~basis label $i \in \{0,1\}^n$, we use the function
\[
  \mathbf{amp}(C_t,i) \;:=\; \langle i \mid C_t \mid 0^n \rangle,
\]
evaluated by contracting the tensor network associated with $C_t$ (e.g., using the algorithm from~\cite{MultiampSim}).
If the operator $U_t$ acts on $n_t$ qubits, then 
\[|\cB_t^{(k)}|=\dim V_t^{(k)}\le 2^{n_t}.\] 
Actually, to find the block $\cB_t^{(k)}$ we use the procedure \BlockDecompose of Algorithm \ref{al:sample},
and the size of the block can be smaller than $2^{n_t}$. For example, the maximum block size of a~controlled 1-qubit gate is 2 independently of the number of controls. In most practical cases, the gates are either classical gates, 1-qubit gates with some number of controls, or an~fSim gate; all of these gates have the maximum block size at most 2 (Fig. \ref{fig:decompose}). 
Thus, for a~circuit composed of such gates, the generation of one sample requires at most $2T$ single amplitude simulator calls.

Let us justify that the connected components used in \textsc{BlockDecompose} indeed define invariant subspaces. Let $A$ be the local gate matrix and build the undirected graph with an edge $\{u,v\}$ whenever $A_{u,v}\ne 0$. If $C$ is a connected component and $u\in C$, $v\notin C$, then $A_{u,v}=0$ and $A_{v,u}=0$, otherwise there would be an edge connecting $C$ to its complement. Hence, after a~permutation of the computational basis, $A$ is block diagonal with blocks indexed by the connected components of this graph. Therefore the span of each component is invariant. If $A$ is unitary, then each block is unitary.

Note that it is also possible to use the multiamplitude simulator from~\cite{MultiampSim} to generate several samples in one run. If we need to generate $K$ samples for a~circuit composed of $T$ standard gates, then it requires $T$ calls of the multiamplitude simulator \cite{MultiampSim}, in each call at most $2K$ amplitudes should be calculated.

When estimating complexity, we should also note that all amplitude calculations are performed on subcircuits of the full circuit. The tensor network contraction cost of any subset of tensors is never larger than that of the entire circuit, since a contraction schedule for any subcircuit can be obtained from the schedule for the full circuit. Hence, the overall sampling task is polynomially reducible to single-amplitude simulation.

We summarize this as a~proposition, given below. Let us
recall that a gate is called \emph{monomial} if its matrix has exactly one nonzero entry in each row and each column (equivalently, it is a product of a diagonal unitary and a permutation matrix). 
For a~circuit $C$ define the following complexity measures:
\begin{itemize}
    \item $\qsize(C)$ is the number of gates in $C$ whose matrices on their full support are \emph{not} monomial. 
    \item $\ell(C)$ is the circuit \emph{locality}, i.e., the maximum number of target qubits acted on by a~single gate (control qubits not counted).
    \item $\CA(C)$ is the complexity of a~single amplitude calculation for  $C$.
    \item $\sCA(C):=\max_{C'\in \mathrm{Sub}(C)}\CA(C')$ is the maximal complexity of a~single amplitude calculation for the circuit $C$ and its subcircuits. Here $\mathrm{Sub}(C)$ denotes the set of all prefix subcircuits of $C$.
    \item $\CS(C)$ is the complexity of generating one sample for  $C$.
    \item $\sCS(C):=\max_{C'\in \mathrm{Sub}(C)}\CS(C')$ is the maximal complexity of one sample generation for subcircuits of circuit $C$.
\end{itemize}

\begin{proposition}
    \label{prop:1}
    For arbitrary quantum circuit $C$
    $$\sCS(C) = O\left(\qsize(C)2^{\ell(C)}\sCA(C)\right).$$
\end{proposition}

\emph{Proof sketch.}
Monomial gates may change the classical state
deterministically but never trigger calls to $\mathbf{amp}(\cdot,\cdot)$.
Only non-monomial gates cause probabilistic branching and hence amplitude
evaluations. Each such gate acts on at most $\ell(C)$ target qubits, so the
block from \textsc{BlockDecompose} has size at most $2^{\ell(C)}$, and updating the state requires at most $2^{\ell(C)}$ evaluations $\bigl|\mathbf{amp}(C_t,i)\bigr|^2$ on a prefix subcircuit $C_t$. Summing over the $\qsize(C)$ non-monomial gates and noting that each call costs at most $\sCA(C)$ yields the bound.

\section{Simulation with noise}

In this section, we show how to adapt the PW simulator to simulate quantum circuits with noisy gates.

\subsection{Noisy simulation algorithm}
It is well known~\cite[Section 8.2.3]{Nielsen-Chuang:2010} that quantum noise can be formally described as an~operator called a~\emph{quantum operation} acting on density matrices $\rho$ using the following operator-sum representation:
$$\mathcal{E}(\rho)=\sum_{i=1}^k E_i\rho E_i^\dagger,\quad \sum_{i=1}^kE_i^\dagger E_i=I.$$
For our purposes it is more convenient to describe the noise operator as a~unitary operator that uses an~ancillary qudit (with $k$ basis states) which can then be measured (e.g., see~\cite[Eq.~(8.42)]{Nielsen-Chuang:2010}). So, the application of a~noise gate is as follows:

\begin{equation}\label{eqn:measure-noise}
    \begin{quantikz}
  \lstick{$\ket{\psi}$} &  \gate[2]{U} \qwbundle{}&\qw \\
  \lstick{$\ket{0}_k$}\qw &   &   \meter{}
\end{quantikz},\hspace{3ex}
\mbox{where }
U=\begin{pmatrix}
    E_1&\cdots\\
    \vdots&\ddots\\
    E_k&\cdots
\end{pmatrix}.
\end{equation}

We want to emphasize that only the first block column of the matrix $U$ is important, since the ancillary qudit is initially in the state $\ket{0}_k$.

Because the ancilla does not interact with the system after $U$, measuring it immediately is equivalent to measuring it later and does not change the outcome statistics for the system; hence we can measure it right away without affecting the system's subsequent evolution.
For different noise operators different ancillary qudits are used. But for our algorithm it is convenient to measure  ancillary qudits immediately because this simplifies the simulation. The measurement value $i$ corresponds exactly to the index of the matrix $E_i$ which we should apply instead of $U$ in our quantum circuit. That is, after we measure the ancillary qudit and get the result of the measurement $i$, we immediately replace the gate $U$ by $E_i$ and update our tensor network:

\begin{equation}\label{eqn:apply-noise}
\begin{quantikz}
  \lstick{$\ket{\psi}$} & \gate[2]{U} \qwbundle{}&\qw&\qw \\
  \lstick{$\ket{0}_k$}\qw &   \qw& \meter{}&\cw \rstick{$i$} 
\end{quantikz}\hspace{1ex}\longrightarrow\hspace{1ex}
\begin{quantikz}
  \lstick{$\ket{\psi}$} &  \gate{E_i} \qwbundle{}&\qw 
\end{quantikz}
\end{equation}

Thus, after adding noise we obtain a~circuit with additional gates $E_i$ in some places (which are not necessarily unitary). Usually, the matrices $E_i$ act on a~single qubit, so we will consider this case as the main case.

Note that when we apply Algorithm \ref{al:sample}, at each step we have the classical state $s$. To apply a noise gate and determine the operator $E_i$ according to \eqref{eqn:apply-noise} and \eqref{eqn:measure-noise}, we usually don't need to add the gate $U$ to the circuit and perform the \UpdateState step. Namely, for most of the noise models (depolarizing, amplitude damping, phase damping) all the matrices $E_i$ have at most one non-zero element in each column and each row. Thus, knowing the state $s$, the measurement probability $i$ of the qudit is equal to the square of the absolute value of the (unique) nonzero element in the column of $E_i$ corresponding to the state $s$: 
\begin{equation}\label{eqn:submonomial-choose-Ei}
    \Pr(i\mid s)=\langle s[q]|E_i^\dagger E_i|s[q]\rangle
\end{equation}
where $q$ is the qubit where the noise gate is applied.

\subsection{Simplifying tensor network via gate propagation}
At each step of the PW simulator, we have a~subcircuit of the initial circuit with additional gates. If we want to calculate transition probabilities for this noisy subcircuit with at most the same complexity as for the circuit without noise, it is enough to build for the noisy circuit a tensor network of the same shape as the tensor network for the initial circuit. 

If the tensor network is built from a quantum circuit in a~straightforward way where an~$m$-qubit gate is converted to a tensor with $2m$ legs, then each additional 1-qubit gate can be fused with one of the neighboring gates.

However, most of the circuits that can be simulated with the proposed algorithm are shallow circuits, like QAOA circuits of low depth and other circuits related to the Ising model. These circuits may contain a~lot of diagonal gates. Each $m$-qubit diagonal gate can be represented as a~tensor with $m$ legs instead of $2m$ for a general gate. This is a~significant simplification. But when we have an~additional non-diagonal $1$-qubit gate between two diagonal gates, and we try to fuse this $1$-qubit gate with an~$m$-qubit diagonal gate, we obtain a~tensor with $(m+1)$ legs, and thus the resulting tensor network is more complex than it was before adding this 1-qubit gate. 

Thus, for a~general noise model the complexity of simulation with noise can be higher than for the circuit without noise if it contains diagonal gates.
However, for all the main noise models: depolarizing noise (and its more general version with unequal probabilities of $X$, $Y$, and $Z$ errors), amplitude damping, and phase damping, all the matrices $E_i$ have at most \emph{one} non-zero entry in each row and each column. Some of these matrices are diagonal and can be immediately fused with neighboring gates without increase in complexity. In the rest of this section we will show how to propagate gates of the form $A=\begin{pmatrix}0&a\\b&0\end{pmatrix}$ through diagonal gates.

The main idea here is that if $P$ is a~permutation matrix, and $D$ is a~diagonal matrix, then the matrix $PDP^\dagger$ is also diagonal with elements on the diagonal permuted according to the permutation $P$. Thus, if the number of qubits on which the permutation gate acts is a~subset of the qubits where the diagonal gate acts, we can propagate a permutation gate through a diagonal gate by the replacement:
$$ PD\mapsto D'P\quad\mbox{where}\quad D'=PDP^\dagger.$$ 
In particular, 1-qubit permutation gates (actually, the only important case is the $X$ gate) can be either propagated (through diagonal gates) or fused (with non-diagonal gates). The gate $A$ can be decomposed as $A=\sigma_X D_A$ where $D_A=\begin{pmatrix}b&0\\0&a\end{pmatrix}$ is the diagonal gate which can be immediately fused with the neighbor gate, and the $X$ gate can be propagated until it is fused with a non-diagonal gate or with a~measurement at the end of the circuit.

 See Appendix~\ref{app:noise-sim} for the technical details and pseudocode of the described algorithm.

\begin{figure*}[!ht]
    \centering
    \includegraphics[width=1\textwidth]{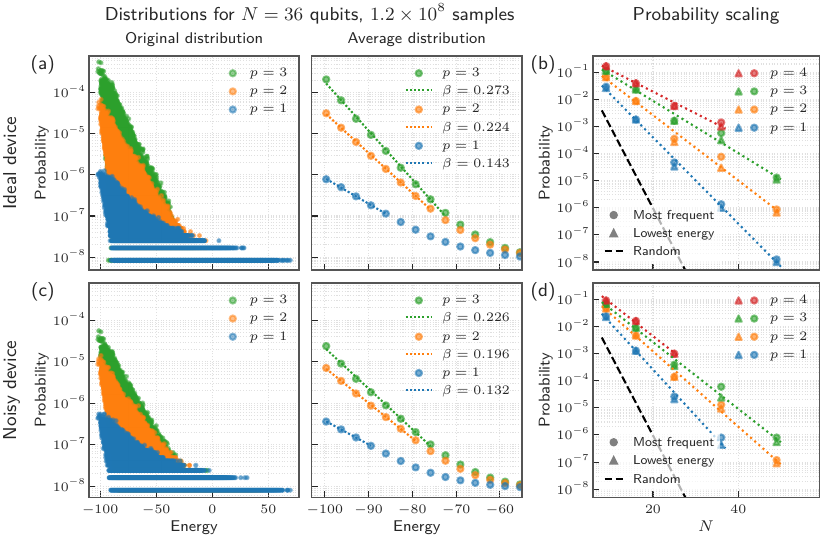} 
    \caption{Sampling results for QAOA circuits obtained with a simulated quantum device. 
    Panels (a,b) correspond to the ideal device; panels (c,d) to the noisy chip. 
    (a,c) Normalized histograms of measurement outcomes from $1.2 \cdot 10^8$ samples on $N=36$ qubits arranged on a rectangular grid. 
    (b,d) Scaling with system size $N$ of (i) the ground state probability and (ii) the probability of the empirically most frequent bitstring in the sampled batch. 
    The dashed black line shows the uniform baseline $2^{-N}$.}        
    \label{fig:distr_and_probability}
\end{figure*}

\section{Experiments}\label{sc:Experiments}

In this section we provide results of numerical experiments to show how our algorithm can be used to obtain meaningful results.

The experiments are conducted on a~server with 512 CPU cores at 2.10 GHz and a~total of 32 TB of RAM, except the experiment in Section~\ref{app:times}, which is performed on another server as stated there.

\subsection{QAOA circuit pseudo-Boltzmann states}
\label{sec:QAOA_pseudo-Boltzmann}
	
Quantum Approximate Optimization Algorithm (QAOA) was proposed in 2014 by Farhi et al. \cite{Farhi2014} as a~variational algorithm to approximate the solution of combinatorial optimization problems. Since then, it has been extensively investigated in various respects and inspired the development of several modifications such as Warm-Start QAOA \cite{Egger2021warmstartingquantum}, ADAPT-QAOA \cite{Zhu2022} and various other ansatze \cite{Hadfield2019}. 
Another direction of research takes advantage of fixed parameters, predicted from some general theoretical assumptions, as in Linear Ramp QAOA \cite{Montanez-Barrera2025} and parameter transfer ~\cite{Basso2022, Shaydulin2023, Sureshbabu2024}, which we adopt for our experiments here. To show the potential of our simulator in obtaining meaningful results in a~realistic scenario, we will show how it can be used to explore the properties of the transferred-parameter QAOA algorithm.

We provide the necessary details of the QAOA algorithm in Appendix \ref{app:QAOA}.

\begin{figure*}[!ht]
    \centering
    \includegraphics[width=1\textwidth]{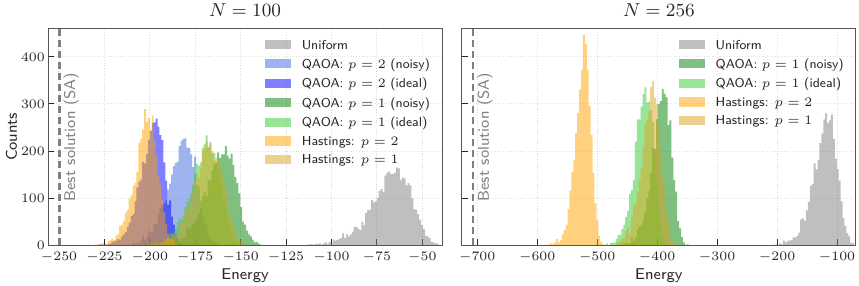} 
    \caption{Experimental results of sampling from QAOA circuit in comparison to classical Hastings algorithm. Each count on the diagram corresponds to the lowest energy out of 100 samples from the corresponding algorithm.}
    \label{fig:Hastings_and_QAOA}
\end{figure*}

In works \cite{Diez-Valle23, Diez-Valle24}, it was shown that the QAOA circuit at depth $p=1$ produces thermal-like distributions for the corresponding Hamiltonians. The probability of a~state in such a~distribution scales as an inverse exponent of its energy: $P(x) \propto e^{-\beta E(x)}$. Here, $\beta$ is the effective inverse temperature, and $E(x)$ is the Ising energy of the state $x$. Original works \cite{Diez-Valle23, Diez-Valle24} only provide simulations for single-layer QAOA circuits of size up to $24$ qubits. The authors suggest that deeper circuits generate Boltzmann distributions with lower effective temperatures. We evaluate their claim, extending the analysis to circuits of up to $p=3$ layers and up to $N=36$ qubits.

Figure \ref{fig:distr_and_probability} (a) shows the probability distribution of states, measured from a~simulated QAOA circuit for the Ising model on a~square lattice of $36$ qubits with normally distributed interaction weights and linear terms. The $y$-axis of the diagram shows the frequency of a~state in the batch of samples, normalized by the total size of the batch. We see that in the lower-energy part, the distribution follows the pseudo-Boltzmann law, as defined in \cite{Diez-Valle23}. It can be seen more clearly on the average plot on the right part of Figure \ref{fig:distr_and_probability} (a). Due to the finite size of the batch, which is small compared to the full space of $2^{36}$ states, the higher energy part represents the density of states in the corresponding energy interval rather than the probability of each particular state and thus does not reflect the Boltzmann law. We can also see that a~deeper circuit indeed provides a~lower effective temperature and a~higher probability of sampling from lower-energy states.

Using the QAOA circuit as an optimization technique, we are interested in obtaining the state with minimal objective value encoded in its Ising energy. For this, in Figure \ref{fig:distr_and_probability} (b), we show the probability of finding the lowest-energy state for different circuit depths and problem sizes of up to 49 qubits. We can see that the probability of finding the ground state of the Ising model with fixed-depth QAOA circuit scales down exponentially, as has been assumed in~\cite{augustino2024}. This trend, although still keeping an~advantage over uniform sampling, may doom the prospects of shallow-depth QAOA to be an~efficient optimization algorithm for large-size problems.

Figure \ref{fig:distr_and_probability} (b) also allows us to discuss the sampling procedure. To provide a~reliable estimate  for the probability, we accumulated the measurement results until the lowest energy state occurred at least 10 times. 
In this setting, depth $p=2$ and $p=3$ circuits were the easiest to model, providing a~balance between simulation complexity and the ground state probability that requires a~smaller number of samples to estimate. For $p=1$, the ground state probability at $N=49$ became so low that we had to generate about $10^9$ samples before the ground state occurred $10$ times. For $p=4$, on the other hand, the complexity of the circuit makes the simulation intractable for $N > 36$. 

Figure \ref{fig:distr_and_probability} (c), (d) shows results similar to Figure  \ref{fig:distr_and_probability} (a), (b), but for sampling from a noisy circuit. Generally, the presence of noise increases the effective temperature and reduces the probability of sampling low-energy states. As a~result, we cannot obtain ground state probability for the 49-qubit circuit of depth $p=1$. Partially, it is due to the lower speed of noisy computation and partially due to the need to obtain a~larger number of samples to reliably calculate ground state probability. Section \ref{app:times} contains information on the time required to generate a~fixed-size batch of samples for different qubit numbers and layout topologies.

For the noisy experiment we selected the following parameters of the simulated chip: single-qubit error probability $0.005$; double-qubit error probability $0.01$; readout error probability $0.05$; single-qubit gate length $30$~$ns$; double-qubit gate length $80$~$ns$; relaxation time $100$~$\mu s$ and dephasing time $50$~$\mu s$. These parameters are arbitrary round numbers near the corresponding values of earlier Sycamore or Zuchongzhi superconducting quantum processors \cite{Arute2019}, \cite{Zuchongzhi_2_1v:2021}.

\subsection{Comparison of QAOA to Hastings algorithm on bigger instances}
\label{sec:QAOA_Hastings}

\begin{figure*}[!ht]
    \centering
    \includegraphics[width=1\textwidth]{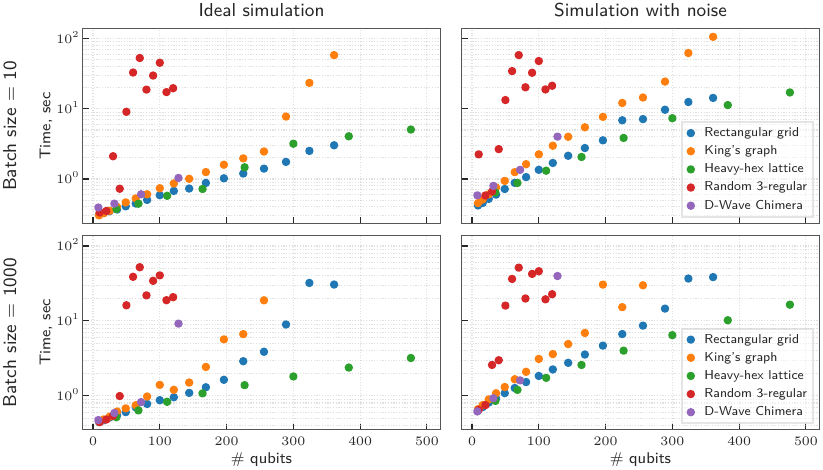} 
    \caption{Sampling times of QAOA circuits with $p=1$, averaged over $10$ independent problem instances for different graph topologies and problem sizes. 
    }
    \label{fig:Times}
\end{figure*}

In~\cite{hastings2019}, Hastings pointed out that the shallow-depth QAOA algorithm is local, in the sense that each layer of a~QAOA circuit updates the degrees of freedom of each qubit based only on its interaction with its immediate neighbors. Hastings argued that a~classical algorithm with similar locality properties should have performance no less than that of QAOA at the corresponding depth:
\begin{displayquote}
``It seems like the QAOA is simply a~way of writing a~quantum circuit which does something similar to a~very simple classical algorithm: initialize spins in some configuration and then update spins depending on those around it. For a~one-step version of the algorithm, the update is usually very simple: if the spins around it apply a~strong enough force to the given spin, then it changes in a~way to increase the objective function.''
\end{displayquote}

Hastings also proposed an example of such a~classical algorithm, see Appendix~\ref{app:Hastings} for details.

Since our simulator works best for sampling from shallow circuits on regular sparse graphs, it is perfectly suited to illustrate this claim. We model shallow-depth QAOA circuits and compare them to several classical algorithms to illustrate their performance on optimization tasks.

We selected two Ising models defined on square lattices of 100 ($10 \times 10$) and 256 ($16 \times 16$) nodes for our experiments. The coupling strengths and linear biases were drawn independently from a~normal distribution. For the $100$-qubit model, we applied QAOA circuits with depths $p = 1$ and $p = 2$, while for the $256$-qubit model, we used only a~single-layer ($p = 1$) circuit. 

Each algorithm takes a~batch of 100 samples and selects the one with the lowest energy as a~solution. We repeat this procedure 4000 times for each algorithm and plot the distribution of the results in Figure \ref{fig:Hastings_and_QAOA}. For reference, we also add the result of simulated annealing (SA), obtained with D-Wave's Neal library \cite{DWave_neal, DWave_neal_git}, and uniform sampling, where each bitstring in the batch is obtained by sampling each bit uniformly at random.

The results of uniform sampling and the SA provide a~natural scale on the energy axis. On this scale, it is clear that the results of the single-step Hastings algorithm are close to depth-$1$ QAOA, and the results of the double-step Hastings are close to depth-$2$ QAOA in the ideal case. It is also clear that quantum noise deteriorates the performance of QAOA, making it inferior to the classical Hastings algorithm at a~similar depth. We note that the purpose of this paper is not to provide a~thorough comparison of the two algorithms, rather to use them as an~illustration of the sampling capabilities of our simulator. So, we used a~simple version of Hastings algorithm, as well as fixed-parameter QAOA circuit (See Appendix \ref{app:Hastings} and \ref{app:QAOA}). Our results show that under reasonable assumptions these algorithms indeed show comparable performance.

\subsection{Sampling times for different topologies and qubit numbers}
\label{app:times}

The experiment in this section is conducted on a~server with $96$ physical CPU cores that supports $2$ threads per core, making a~total of $192$ available threads. The base CPU frequency is $2.3$~GHz, with a~maximal boost frequency of $3.3$~GHz, and the server has $1.5$~TB of memory.

In Figure~\ref{fig:Times} we show the average time to generate a~batch of samples of fixed size from simulated ideal and noisy devices of different qubit numbers and layout topologies. Some of the topologies correspond to the layouts of real quantum devices. The rectangular grid layout is used in Google's processors~\cite{Arute2019, Acharya2024}, as well as similar devices from USTC \cite{Zuchongzhi_2_1v:2021, gao2024establishingnewbenchmarkquantum} and Zhejiang University \cite{Zhu2026}. The heavy-hex lattice is typical for IBM's processors \cite{IBM_Heavy-hex}, while the King's graph topology is relevant to some experiments with cold atoms \cite{Kim2024}. We also add the Chimera topology, native for earlier models of D-Wave's quantum annealers \cite{DWave_Chimera}. Although those devices do not support execution of the QAOA algorithm, the Chimera graph is a~good example of regular, though non-planar topology, which allows us to demonstrate the limits of our simulator. Finally, we add random 3-regular graphs, which are important in theoretical studies of QAOA \cite{Farhi2014}, and also represent a~more difficult, though still tractable case for our simulator. We simulate a~QAOA circuit of depth $p=1$ for all experiments and take the average over $10$ independent runs, each run on a~separate random instance of the Ising model with Gaussian weights.

We can clearly see that sparser topologies are simulated faster than denser ones, and that regular graphs are easier to simulate than random graphs.

\section{Conclusions}

In this work, we introduced a~novel classical sampling algorithm that combines the frameworks from~\cite{pilot-wave-sim-2, bravyiHowSimulateQuantum2022} with tensor network contraction techniques, and showed that this algorithm can be naturally extended to incorporate realistic noise models, enhancing its relevance for studying contemporary quantum devices. This new approach is particularly well-suited for efficiently producing samples from the output of structured, non-random quantum circuits, a~regime where methods like rejection sampling often struggle.

We emphasize that the method should be understood as an exact sampling layer on top of a~tensor-network amplitude backend. It does not reduce the cost of a~single tensor-network contraction, and its practical advantage is expected mainly for shallow, sparse, structured circuits where the required prefix amplitudes remain tractable. In the noisy case, the low-overhead implementation relies on noise channels admitting a local-submonomial Kraus representation, such as the depolarizing and amplitude-damping channels used in our experiments.

Through extensive numerical experiments, our simulator effectively handles QAOA circuits of up to $476$ qubits at depth $p=1$, $100$ qubits at depth $p=2$ and $49$ qubits at depth $p=3$, significantly surpassing the capabilities of existing classical methods such as rejection sampling, especially for circuits with highly non-uniform output distributions.

We specifically explored the Quantum Approximate Optimization Algorithm (QAOA), highlighting several key insights. Our findings confirm and extend previous observations that QAOA circuits produce pseudo-Boltzmann distributions, demonstrating that deeper circuits lead to lower effective temperatures and thus enhance sampling from lower-energy states. Importantly, we quantified how the probability of finding the ground state decreases exponentially with problem size, underscoring limitations of shallow-depth QAOA for practical optimization tasks at large scales.

Our comparative analysis with Hastings' classical local-update algorithm revealed that classical algorithms at comparable depth can perform similarly to QAOA, challenging the utility of current shallow-depth quantum approaches for optimization problems. Additionally, by integrating realistic quantum noise models into our simulations, we 
assessed the detrimental impact of noise on the effectiveness of quantum algorithms, showing potential of our simulator as a tool for benchmarking near-term quantum hardware. Beyond QAOA, the same structural conditions suggest applicability to other shallow, locally-interacting circuit families such as hardware-efficient VQE ansätze and Trotterized simulations of local Hamiltonians, where bounded gate locality and low-dimensional block structure preserve the constant per-gate cost of the PW simulation framework.

Overall, our simulator not only offers an efficient classical approach to exploring the behavior of structured quantum circuits but also serves as a powerful tool for guiding the development and benchmarking of quantum algorithms on current and future noisy intermediate-scale quantum (NISQ) devices.

\section*{Code and data availability}

An~open-source implementation of the Pilot-Wave simulator is available at \url{https://github.com/KalachevGleb/pilot-wave-simulator}. The repository contains documentation, scripts for the numerical experiments, input instances, parameters, random seeds, and processed data used to generate the figures. Very large raw streams of samples are not stored in the repository because of their size, but they can be regenerated using the provided scripts.

\section*{Author contribution}
G.K and P.P designed the core algorithm and developed the code. Z.Z and P.M designed and performed numerical experiments. M-H.Y. coordinated the work. All authors contributed to discussion of the results and text writing.

Authors used generative AI tools to refine the text and prototype the code during the revision.

\section*{Acknowledgment}
This work is supported by the Shenzhen International Quantum Academy (Grant No.SIQA2024KFKT01).

\bibliographystyle{unsrturl} 
\bibliography{simulation}

\onecolumngrid

\clearpage
\appendix
\section{\textbf{Noisy simulation details} }\label{app:noise-sim}

In this appendix, we give a~detailed description of our sampling algorithm from the output of a~noisy quantum circuit $C'$, obtained from the circuit $C$ by inserting noise gates according to the model.
A quantum circuit is represented as a list of gates, each denoted by a pair $(A,q)$, where $A$ is a $2^m\times 2^m$ matrix, and $q=(q_1,\dots,q_m)$ is a tuple of $m$ qubit indices $q_i\in \{1,\dots,n\}$, which we identify with the corresponding qubits. The notation $X_q:=(\sigma_X,(q))$ indicates the Pauli-X operator  acting on qubit  $q\in \{1,\dots,n\}$ in the quantum circuit.

\begin{algorithm}[ht]
{
\caption{Circuit simplification via gate fusion and propagation}\label{alg:noise-sim}
  \SetKwProg{Function}{Function}{:}{}
  \SetKwFunction{SimulateCircuit}{NoisySample}
  \SetKwFunction{CanFuse}{CanFuse}
  \SetKwFunction{PropagateX}{PropagateX}
  \SetKwFunction{AddGate}{AddGate}

  \LinesNumbered 
  \KwIn{Noisy circuit $C'$}
  \KwOut{Sample generated according to the output distribution of circuit $C'$}

  \Function{\SimulateCircuit{$C'$}}{
    Let $C'' \leftarrow \text{the empty circuit}$\;
    Let $s:=\mathbf{0}\in\{0,1\}^n$\;
    Let $x:=\mathbf{0}\in\{0,1\}^n$\;
    \ForEach{gate $G$ in $C'$ in execution order}{
      \uIf{$G$ is a~unitary gate}{
         $C'',x,s\gets$ \AddGate{$C'', G, x, s$}\; \label{alg-ln:normal-add-gate}
      }
      \Else{\tcp{$G$ is a~noise gate}
        Apply the unitary representation $U_G$ of $G$ after the circuit $C''$ with an~ancilla qudit as shown in Eq. \eqref{eqn:measure-noise}\; \label{alg-ln:measure-noise}
        Determine the operator $E$ as shown in Eq. \eqref{eqn:apply-noise}\;\label{alg-ln:apply-noise}
        $L\gets$ the decomposition of $E$ as a~tensor product of smaller gates\;
        \ForEach{$(M,q)\in L$}{
            \uIf{$q$ is a single qubit and $M\sigma_X$ is diagonal}{
                $x[q]\gets 1-x[q]$\;
                $M\gets M\sigma_X$\; \label{alg-ln:antidiag2diag}
            }
            $C'',x,s\gets$ \AddGate{$C'', (M,q), x, s$}\; \label{alg-ln:noise-add-gate}
        }
      }
    }
    \Return{$(s+x)\mod 2$}\;
  }
  \Function{\AddGate{$C,G,x,s$}}{
     $G'\gets \PropagateX{x,G}$\;
     Add $G'$ to list of gates in $C$\; \label{alg-ln:add-gate-real}
     \uIf{$G'$ is not diagonal}{
        $s\gets \mathrm{UpdateState}(C,s)$\; \label{alg-ln:update-state}
     }
     \Return{$C,x,s$}
  }

  \Function{\PropagateX{$x,G$}}{
      \ForEach{$q\in \mathrm{qubits}(G)$}{
        \uIf{$x[q]=1$}{
            $G\gets X_q\circ G \circ X_q$\;
        }
      }
      \Return{$G$}
  }
}
\end{algorithm}

To efficiently simulate a quantum circuit with stochastic noise using the proposed method, we need to ensure that the complexity of computing a~single amplitude of the \emph{noisy circuit}~$C'$ (i.e., the circuit~$C$ with additional noise gates) remains the same as that of the original circuit~$C$, under certain conditions. Since our method is based on tensor network contraction, it is enough to show that a single amplitude for $C'$ can be calculated by a~tensor network of the same shape as for the circuit $C$.

Initially, an~$m$-qubit gate $G$ is naturally represented as a~tensor $T_G$ with $2m$ legs, where each qubit corresponds to two legs. Then, for each pair of legs $i,j$ corresponding to qubit $q$, if all entries of $T_G$ with indices $i\ne j$ are zero, then we identify legs $i$ and $j$; in this case, we say that a~gate $G$ is \emph{diagonal on qubit} $q$. Gate $G$ is \emph{diagonal} if it is diagonal on all qubits, or, equivalently, its matrix is diagonal. 
For example, for a~diagonal gate $G$, all output legs of $T_G$ will be identified with corresponding input legs. 
It is important to note that any one-qubit diagonal gate, such as $Z$, can be represented by a~tensor with one leg.

With this simplification, for example, a~QAOA circuit of depth $1$ for the Max-Cut problem on the graph $\Gamma$ corresponds to the tensor network that is isomorphic to graph $\Gamma$ where the edges correspond to tensors and the vertices correspond to the~legs. The tensor network of the QAOA circuit of depth $p$ is isomorphic to the Cartesian product of the graph $\Gamma$ and the chain with $p$ vertices.

Now let us formulate the sufficient condition when the noisy circuit can be simulated without overhead (with the same tensor contraction complexity). We say that a~$2^m\times 2^m$ matrix $M$ is \emph{local-submonomial} (\emph{LSM}) if it is simultaneously local and submonomial, i.e.,  $M=\bigotimes_{i=1}^mM_i$ where $M_i$ are $2\times 2$ matrices and for each $M_i$ its rows and columns contain \emph{at most one} nonzero entry. We say that a~quantum operation $\mathcal{E}$ is \emph{local-submonomial} (\emph{LSM})  if it admits a~Kraus
representation $\mathcal{E}(\rho)=\sum_i E_i\rho E_i^\dagger$ in which
all $E_i$ are LSM matrices.

\begin{proposition}
    If noise can be represented as LSM quantum operations, then the tensor contraction complexity in Algorithm \ref{alg:noise-sim} is not higher than the tensor contraction complexity in Algorithm \ref{al:sample} with an overhead of $O(\mbox{number of noisy gates})$.
\end{proposition}

\begin{figure}
    \centering
    \subfloat[non-diagonal case (fusion)]{\label{fig:x-fuse}
    \begin{tikzpicture}[scale=1.5]
        \foreach \x in {(-0.2,0), (0.8,0)}{ \fill \x circle[radius=1.5pt];  }
        \draw (1,0.2) rectangle node{$G$} (1.4,-1.2);
        \draw (-0.2,0)
        --(0.2,0);
        \draw (0.2,-0.2) rectangle node{$A$} (0.6,0.2);
        \draw (0.6,0)--(0.8,0) node[below] {$i$} -- (1,0); 
        \foreach \y in {0,-0.4,-1} {
        \draw (1.4,\y)--++(0.4,0);
        }
        \draw (1,-0.4)--(-0.2,-0.4);
        \draw (1,-1)--(-0.2,-1);
        \node at (-0.1,-0.6) {$\vdots$};
        \node at (1.7,-0.6) {$\vdots$};
        \draw[blue, thick] (0.1,0.3) rectangle(1.5,-1.3);
    \end{tikzpicture}
    }
    \hspace{0.5cm}
    \subfloat[diagonal case, $G'_{k,*}=G_{1-k,*}$]{\label{fig:x-propagate}
    \begin{tikzpicture}[scale=1.5]
        \foreach \x in {(0.0,0), (1.0,0), (2.4,0), (3.4,0)}{ \fill \x circle[radius=1.5pt];  }
        \draw (-0.2,0) -- (0,0) node[above]{$i\vphantom{j}$} --(0.2,0);
        \draw (0.2,-0.2) rectangle node{$X$} (0.6,0.2);
        \draw (0.6,0)--(1.0,0)node[above]{$j$}--(1.0,-0.4);
         \draw (1.0,0)--(1.4,0);
        \draw (-0.2,-0.4) -- (1.4,-0.4);
        \draw (-0.2,-1.0) -- (1.4,-1.0);
        \node at (-0.1,-0.6) {$\vdots$};
        \node at (1.3,-0.6) {$\vdots$};
        \draw[fill=white] (0.8,-1.1) rectangle node{$G$} (1.2,-0.3);
        \draw (2.4,0)--++(0,-0.4);
        \draw (3.2,-0.2) rectangle node{$X$} (2.8,0.2);
        \draw (2.8,0)--++(-0.4,0)node[above]{$i\vphantom{j}$} -- ++(-0.4,0);
        \draw (3.2,0)--(3.4,0) node[above] {$j$} -- (3.6,0);
        \draw (2.0,-0.4) -- (3.6,-0.4);
        \draw (2.0,-1.0) -- (3.6,-1.0);
        \node at (2.1,-0.6) {$\vdots$};
        \node at (3.5,-0.6) {$\vdots$};
        \draw[fill=white] (2.6,-1.1) rectangle node{$G'$} (2.2,-0.3);
        \draw[-stealth, blue, thick] (1.5,-0.7)--(1.9,-0.7);
    \end{tikzpicture}
    }
    \caption{Fusion and propagation of an~$X$ gate in a~quantum circuit. }
    \label{fig:my_label}
\end{figure}
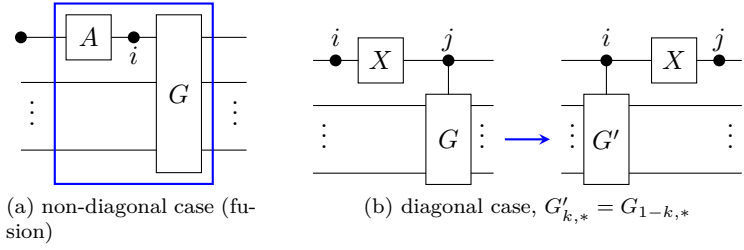

The overall sampling algorithm is described in Func. \SimulateCircuit in Algorithm \ref{alg:noise-sim}. It is not hard to see that the obtained tensor network during the execution of the algorithm at each step corresponds to the list of gates in circuit $C''$. Gates are added to $C''$ in lines \ref{alg-ln:normal-add-gate} and \ref{alg-ln:noise-add-gate}. Line \ref{alg-ln:normal-add-gate} adds to $C''$ circuit gates which are the same as in the simulation without noise. Additional gates are added in line \ref{alg-ln:noise-add-gate} where we consider noise gates. For local-submonomial noise, the tensor product decomposition of $E$ consists of single-qubit operators which are either diagonal (can be immediately fused) or antidiagonal (reduced to the diagonal case in line \ref{alg-ln:antidiag2diag}).
The main idea is that the $X$ gate can be propagated through any other gate without changing the shape of the corresponding tensor (Func. \PropagateX, see also Fig. \ref{fig:x-propagate}). Thus, the matrix $M$ in line \ref{alg-ln:noise-add-gate} is always diagonal, and $q$ is always a single qubit, so the added gate is always a 1-qubit diagonal gate. Hence, \UpdateState in line \ref{alg-ln:update-state} is never called from \AddGate called from line \ref{alg-ln:noise-add-gate}. In addition, the circuit $C''$ for local-submonomial noise differs from the noiseless prefix subcircuit of $C$ only by some number of diagonal 1-qubit gates. 
Calculation of an amplitude of $C''$ via tensor contraction consists of 2 steps: (i) fusing one-leg tensors corresponding to 1-qubit diagonal gates with tensors sharing this leg; (ii) contracting the obtained tensor network using standard techniques.
Thus, for local-submonomial noise, the tensor network obtained at step (ii) will have the same graph as for the noiseless circuit, and the same contraction order can be used to contract it with the same complexity. 
Thus, the complexity of tensor contraction in noisy simulation is equivalent to the complexity of tensor contraction in simulation without noise when \UpdateState is executed.
An~additional potential place where tensor contraction is performed at lines \ref{alg-ln:measure-noise} and \ref{alg-ln:apply-noise} of Func. \SimulateCircuit, where virtually an additional gate and ancilla qudit are added to the circuit. However, as it was noted, for local-submonomial noise operators, these steps can be done using \eqref{eqn:submonomial-choose-Ei} without adding gates and performing a real tensor contraction.

\newpage

\section{Ising model}

The Ising model is a~paradigmatic mathematical framework in statistical mechanics that represents a~system of $N$ interacting spins through its Hamiltonian
\begin{equation}
    \label{isingH}
    H = \sum_{i, j = 1}^N J_{ij}\sigma_i \sigma_j + \sum_{i = 1}^N h_i \sigma_i,
\end{equation}
where $J_{ij} = J_{ji}$ is half the coupling strength between spins $i$ and $j$, while $h_i$ is the parameter characterizing the longitudinal field. $J_{ii} =0$. Variables $\sigma_i$ take discrete values $\pm1$, and can be represented by the Pauli $Z$ operator in a~quantum circuit. 

It is proven that finding the ground state of Ising Hamiltonian is an NP-hard problem \cite{Barahona1982}, which makes it important for combinatorial optimization research. At the same time, sampling from the corresponding Boltzmann distribution  is an~important task in statistical physics and adjacent fields.

\section{Hastings algorithm}
\label{app:Hastings}

In~\cite{hastings2019}, a~family of local approximate optimization algorithms was proposed and claimed to be competitive with QAOA at similar depth. Define a set of vectors $\vec{v}_t$ for each $t = 0,\dots, p$, where $\vec{v}_0$ is an initial solution that can be initialized randomly. In our experiment, it is a~vector consisting of values $\pm 1$ taken randomly with equal probability. Then at each step the vector is updated according to the following rule:
\begin{equation}
    \vec{v}_{t+1} = g_t \left( \vec{v}_t + c_t \cdot \vec{F_t} \right),
\end{equation}
where $\vec{F_t}$ is a ``force vector'', $c_t$ are hyperparameters, and $g_t$ are some real functions. Each entry of the force vector is defined as half the difference between the value of the objective function with the corresponding variable set to $+1$ and $-1$, and the other variables substituted with the corresponding values of $\vec{v}_t$.  Thus, in our case with the Ising model (\ref{isingH}) as the objective, the force vector is defined as follows:
\begin{equation}
    \vec{F_t} = 2 J \cdot \vec{v}_t + \vec{h}.
\end{equation}

In our simulation, we used $c_t = 0.3$ for all $t$ and both problem sizes. The transformation $g_t$ is chosen to be the identity. The sign of the parameters $c_t$ determines whether the algorithm will maximize or minimize the objective function. After $p$ iterations, the values of spins in the Ising model are computed as signs of corresponding entries in $\vec{v}$: $\sigma_i = \mathrm{sign}(\vec{v}_{p,i})$.

In the present work, we used only the single-step and double-step Hastings algorithms to provide a~direct comparison with single-layer and double-layer QAOA.

\section{QAOA algorithm}
\label{app:QAOA}

The QAOA algorithm \cite{Farhi2014}, \cite{BLEKOS20241} is implemented in the following steps:

\begin{enumerate}
    \item Define cost and mixer Hamiltonians $H_C$ and $H_M$ that do not commute with each other. Cost Hamiltonian corresponds to Ising energy and is given by the following formula:
    \begin{equation}        
        H_C = \sum_{i, j} J_{i, j} Z_i Z_j + \sum_i h_i Z_i,
    \end{equation}
    where $J_{i, j}$ and $h_i$ are Ising coupling matrix and linear shift fields, $Z_i$ are Pauli Z operators. Mixing Hamiltonian is a~sum of Pauli X operators acting on each qubit: $H_M = \sum_i X_i$.
    
    \item Initialize the circuit in equal superposition of all possible states by applying Hadamard gate to each qubit:
    \begin{equation}
        \ket{s} = \ket{+}^{\otimes n} = H^{\otimes n} \ket{0}^{\otimes n}.
    \end{equation}

    \item Construct the circuit ansatz by defining the cost and mixer unitary operators:

    \begin{align}
        \hat{U}_C(\gamma) & = e^{-i \gamma H_C} \\
        \hat{U}_M(\beta) & = e^{-i \beta H_M}.
    \end{align}

    Here, $\gamma$ and $\beta$ are variational parameters of the circuit. We use the rotation convention $R_P(\theta) := e^{-i \theta P / 2}$ for any Pauli product $P$. Each term in cost and mixer operator can be implemented with rotation gates as follows:

    \begin{align}
        & e^{-i \gamma J_{i, j} Z_i Z_j} = R_{Z_i Z_j} (2 \gamma J_{i, j}), \\
        & e^{-i \gamma h_i Z_i} = R_{Z_i} (2 \gamma h_i), \\
        & e^{-i \beta X_i} = R_{X_i} (2 \beta),
    \end{align}

    which allows us to implement $\hat{U}_C(\gamma)$ as a~circuit layer consisting of single and double-qubit Z-rotation gates, and $\hat{U}_M(\beta)$ as a~layer of single qubit X-rotation gates.    

    \item With a~given circuit depth $p \ge 1$, define parameters $\boldsymbol{\gamma} = (\gamma_1, \dots, \gamma_p)$; $\boldsymbol{\beta} = (\beta_1, \dots, \beta_p)$ and apply the full circuit to obtain the final quantum state:

    \begin{equation}
        \ket{\Psi_p(\boldsymbol{\gamma}, \boldsymbol{\beta})} = \hat{U}_M(\beta_p) \hat{U}_C(\gamma_p)  \dots  \hat{U}_M(\beta_1) \hat{U}_C(\gamma_1) \ket{s}.
    \end{equation}
    
\end{enumerate}

Original QAOA is a~variational algorithm, which means that after construction of the quantum state $\ket{\Psi_p(\boldsymbol{\gamma}, \boldsymbol{\beta})}$, the expectation value of the cost Hamiltonian $H_C$ with respect to this state is being optimized in a~classical loop by repeating the procedure and adjusting the parameters $\boldsymbol{\gamma}$ and $\boldsymbol{\beta}$. For the present work, since our goal is to show sampling capabilities for a~given quantum circuit, we used a~fixed set of parameters according to the transfer technique from~\cite{Basso2022, Shaydulin2023} and the scaling for $\boldsymbol{\gamma}$ from~\cite{Sureshbabu2024}.

\section{Applicability beyond QAOA under structural constraints}

The PW simulator applies to circuits satisfying the structural conditions in Section \ref{sec:PW_sim}: bounded locality $\ell(C)=O(1)$, low-dimensional gate decompositions, and efficient tensor-network evaluation of prefix amplitudes (Proposition \ref{prop:1}). Under these constraints, sampling cost is governed by constant-size block updates independent of system size, with QAOA serving as a representative benchmark rather than a unique case.

A direct extension is hardware-efficient VQE ansätze on sparse connectivity graphs, where alternating single-qubit rotations and local entanglers preserve bounded block structure (typically $\le 4$). As long as circuit depth remains shallow, the same Markov update rule and amplitude oracle remain applicable without modification.

Similarly, Trotterized simulations of local Hamiltonians fall into the same regime. In the transverse-field Ising model, $R_{zz}$ and $R_x$ layers yield identical block structure to QAOA. In the Heisenberg model, $XX+YY$ interactions act on a two-dimensional subspace ${|01\rangle,|10\rangle}$ while diagonal terms remain free. For the Hubbard model under Jordan–Wigner mapping, interaction terms are diagonal and hopping terms reduce to the same constant-size $(XX+YY)$ blocks dressed with diagonal $Z$-strings, which do not increase complexity.

These cases share the same essential property: all non-monomial operations admit bounded block size, ensuring constant per-gate sampling cost and preserving efficient contraction. Circuits with large non-local entangling structure or unbounded block decomposition fall outside this regime.

\end{document}